\pdfoutput=1
\documentclass{article}

\usepackage[dblblindworkshop, final]{neurips_2026}
\workshoptitle{PALM: Personalized, Aligned, Long-Term Memory for AI Systems}

\usepackage[T1]{fontenc}
\usepackage[utf8]{inputenc}
\usepackage{booktabs}
\usepackage{graphicx}
\usepackage{multirow}
\usepackage{amsmath,amssymb}
\usepackage{array}
\usepackage{xcolor}
\usepackage{microtype}
\usepackage[hidelinks]{hyperref}
\usepackage[capitalise,noabbrev]{cleveref}

\newcommand{\pass}{\textsc{pass}}
\newcommand{\fail}{\textbf{\textsc{fail}}}
\newcommand{\incon}{\textsc{inconclusive}}
\newcommand{\refuted}{\textsc{refuted}}
\newcommand{\best}[1]{\textbf{#1}}
\newcommand{\ci}[2]{{\footnotesize[#1,\,#2]}}
\newcommand{\rhotrust}{\rho}

\title{Surfacing Disagreement Beats Resolving It:\\
A Gated Ablation Study of Provenance-Weighted\\ Memory for Multi-Agent Debate}

\author{%
  Ziaullah Khan$^{1}$ \quad Abdul Rehman Khalid$^{1}$ \quad Shrestha Nischal Lal$^{2}$ \\
  \textbf{Nheng Vanchhay$^{1}$ \quad Md Ariful Islam Mozumder$^{1}$ \quad
    Hee-Cheol Kim$^{1}$\thanks{Corresponding author.}} \\[4pt]
  $^{1}$Department of Digital Anti-Aging Healthcare, Inje University, Gimhae, Republic of Korea \\
  $^{2}$Department of Computer Engineering, Inje University, Gimhae, Republic of Korea \\[4pt]
  \texttt{\{ziakhan, arkhalid, nhengvanchhay, heeki\}@live.inje.ac.kr} \\
  \texttt{nischal@oasis.inje.ac.kr}, \texttt{arif@inje.ac.kr}
}

\begin{document}
\maketitle

\begin{abstract}
Shared memory for multi-agent debate is usually built to \emph{resolve} inconsistency:
retrieval returns one winning fact and the disagreement disappears before any agent sees
it. We ask whether that is correct. In DebateGraph, four backends share one contract
(\texttt{read}/\texttt{write}/\texttt{snapshot}), so five mechanisms from the
agent-memory literature can be removed independently under a fixed budget, each
gated on a pre-declared threshold. Across 39 runs one survives:
\textbf{preserving contradictions and passing the
ranked conflict set to the debate, rather than collapsing it to a winner, improves
F1 by $+0.340$ (95\% CI $[+0.267,+0.414]$) and cuts attack success from $22$ to $7$ of
$122$ poisoned items (McNemar exact $p{=}0.0003$)}, at equal cost. Reaching it meant
fixing a bottleneck in our harness: conflict sets are built only over retrieved edges, and
our pre-declared cap left three quarters of planted contradictions unretrievable, first
measuring the mechanism at $+0.084$. The effect replicates on two retrieval substrates
and directionally on HotpotQA documents. No other mechanism survives, and two are worse
than inert: once contradictions reach the prompt, a write trust-gate and the router's
provenance features each \emph{cost} accuracy ($-0.075$ and $-0.096$ F1). The gate fails
structurally: poisoned objects are always entities the graph already contains, against
$53\%$ of gold, so groundedness scores the lie \emph{above} a truthful first mention. We
release the harness, attack set, and logs, and argue that silently reconciling memory is
the assumption worth abandoning.
\end{abstract}

\section{Introduction}

As debates lengthen and span sessions, agents need shared memory, and recent work supplies
it: episodic buffers with learned consolidation~\citep{packer2023memgpt,chhikara2025mem0},
temporal knowledge graphs~\citep{rasmussen2025zep}, hierarchical multi-agent
stores~\citep{zhang2025gmemory}, and graph-structured
retrieval~\citep{gutierrez2024hipporag,edge2024graphrag}.

These systems share a design reflex that is rarely stated as a hypothesis: \emph{memory
should be consistent}. When two stored facts conflict, the store reconciles them (by
recency, by a confidence score, by an LLM-adjudicated merge), and retrieval hands the debate
a single clean answer. The disagreement is resolved by the storage layer and never reaches
the agents. Conflict itself is not unexamined; a recent line of work builds taxonomies,
benchmarks and methods for conflicting retrieved evidence. But as \cref{sec:related} sets
out, that work overwhelmingly resolves the conflict too, only earlier or elsewhere. What is
untested is the division of labour: whether the storage layer should be the component that
decides.

This is a strange division. A debate system is, by construction, machinery for adjudicating
competing claims with explicit reasoning. Collapsing a conflict before the debate begins
spends the storage layer's weakest evidence (a heuristic tie-break) to pre-empt the
system's strongest capability. It is also brittle where it matters for shared memory: if an
adversary can write, and the tie-break is recency, the adversary chooses the winner by
choosing a timestamp.

We test the reflex directly, in a service where the read path, write path, consistency model
and retrieval policy are separable, and where suppressing versus surfacing a contradiction
is a one-flag ablation with everything else fixed. The contribution is not a new state of
the art (dumping the whole conversation into the prompt remains the most accurate
configuration by a wide margin) but a controlled measurement of \emph{which} proposed
mechanisms earn their place, under a protocol that fixes each threshold before the run.

\textbf{Contributions.} (i) A memory service in which five literature mechanisms are
independent switches behind one contract. (ii) DebateMem-Conflict, a deterministic
provenance-labelled contradiction-injection set. (iii) A gated ablation over 39 runs in
which one mechanism survives, two are measurably harmful, and the harness bottleneck that
initially hid the effect by a factor of eight is reported rather than quietly corrected.

\section{Related Work}
\label{sec:related}

\textbf{Debate and memory.} \citet{du2024debate} establish multi-agent debate;
\citet{liang2024divergent} and \citet{wang2023selfconsistency} vary how agreement is
reached. Memory systems supply the shared store~\citep{packer2023memgpt,chhikara2025mem0,
rasmussen2025zep,zhang2025gmemory}, and graph retrieval supplies its structured
form~\citep{gutierrez2024hipporag,edge2024graphrag}. None isolates what the store should do
with a contradiction.

\textbf{Conflicting evidence, and what is done with it.} A parallel line treats conflict as
a first-class retrieval problem. \citet{cattan2025dragged} give a taxonomy of
knowledge-conflict types with an annotated benchmark, showing models resolve them poorly
unless told the type; \citet{wang2025madamrag} pair a conflicting-evidence dataset with a
multi-agent method giving each retrieved document its own agent. Inside debate,
\citet{tian2026madmm} improve robustness by \emph{masking} erroneous memories between
rounds, and \citet{meng2026equimem} calibrate what may be written to shared memory by
scoring each update against global memory state.

We take this work as the premise rather than the gap. Three of the four still resolve before
the reasoner sees the disagreement (by classifying it, discarding it, or admitting only
clean writes), and the fourth distributes it across agents. None isolates our question:
holding the debate, budget and retrieved set fixed, should the \emph{storage layer}
adjudicate at all when its consumer already does? \Cref{sec:gate} is the sharpest contact
point: our write-time gate scores groundedness per candidate and fails structurally, and our
diagnosis (that the discriminating signal is \emph{relative}, candidate against incumbent
on the same key) is an independent argument for evaluating writes against global memory
state as \citet{meng2026equimem} do.

\textbf{Poisoning and trust.} PoisonedRAG~\citep{zou2024poisonedrag} corrupts corpora;
AgentPoison~\citep{chen2024agentpoison} backdoors agent memory. Our $\rhotrust$ descends
from knowledge fusion~\citep{dong2014knowledgevault} and the functionality filter from
AMIE~\citep{galarraga2013amie}; round-level consistency is textbook
MVCC~\citep{bernstein1987concurrency}.

\section{Problem Statement}
\label{sec:problem}

A debate instance is $(q,\mathcal{S},B)$: a question, a memory scope, and a hard token budget
for retrieved context. $N$ agents deliberate for $R$ rounds; a judge emits $\hat{a}$, scored
by token-F1 against gold. Memory holds \emph{episodes} and \emph{edges}; an edge carries full
provenance and lifecycle,
\begin{equation}
\epsilon = \langle h,r,t,\ \text{writer},\ t_{\text{ing}},\ t_{\text{valid}},\ t_{\text{inval}},\ \rhotrust,\ \sigma,\ \text{supp}\rangle,
\label{eq:edge}
\end{equation}
with $\rhotrust\in[0,1]$ its trust and $\sigma$ its source type. Edges are never deleted:
supersession sets $t_{\text{inval}}$ and both versions persist.

\textbf{Conflict.} Let $\mathcal{E}_{h,r}$ be all edges sharing subject $h$ and relation $r$.
A \emph{conflict} exists when $\mathcal{E}_{h,r}$ holds at least two distinct objects. The
question this paper asks is what \textsc{read} should do with it:
\begin{equation}
\textbf{collapse: } \mathcal{C} \ni \arg\max_{\epsilon\in\mathcal{E}_{h,r}} \pi(\epsilon)
\qquad\text{versus}\qquad
\textbf{surface: } \mathcal{C} \ni (\mathcal{E}_{h,r},\pi),
\label{eq:policy}
\end{equation}
for a ranking $\pi$. Collapse is what current systems do. Surface returns the whole ranked
set with provenance attached and leaves adjudication to the agents.

\textbf{Threat model.} An adversary may write to shared memory but cannot modify existing
edges, read agent state, or alter prompts. It knows the corpus vocabulary and controls the
timestamp of its own writes. This is the realistic case for a compromised agent, and the
model under which recency-based tie-breaks fail.

\section{Method}
\label{sec:method}

\begin{figure}[t]
\centering
\includegraphics[width=\textwidth]{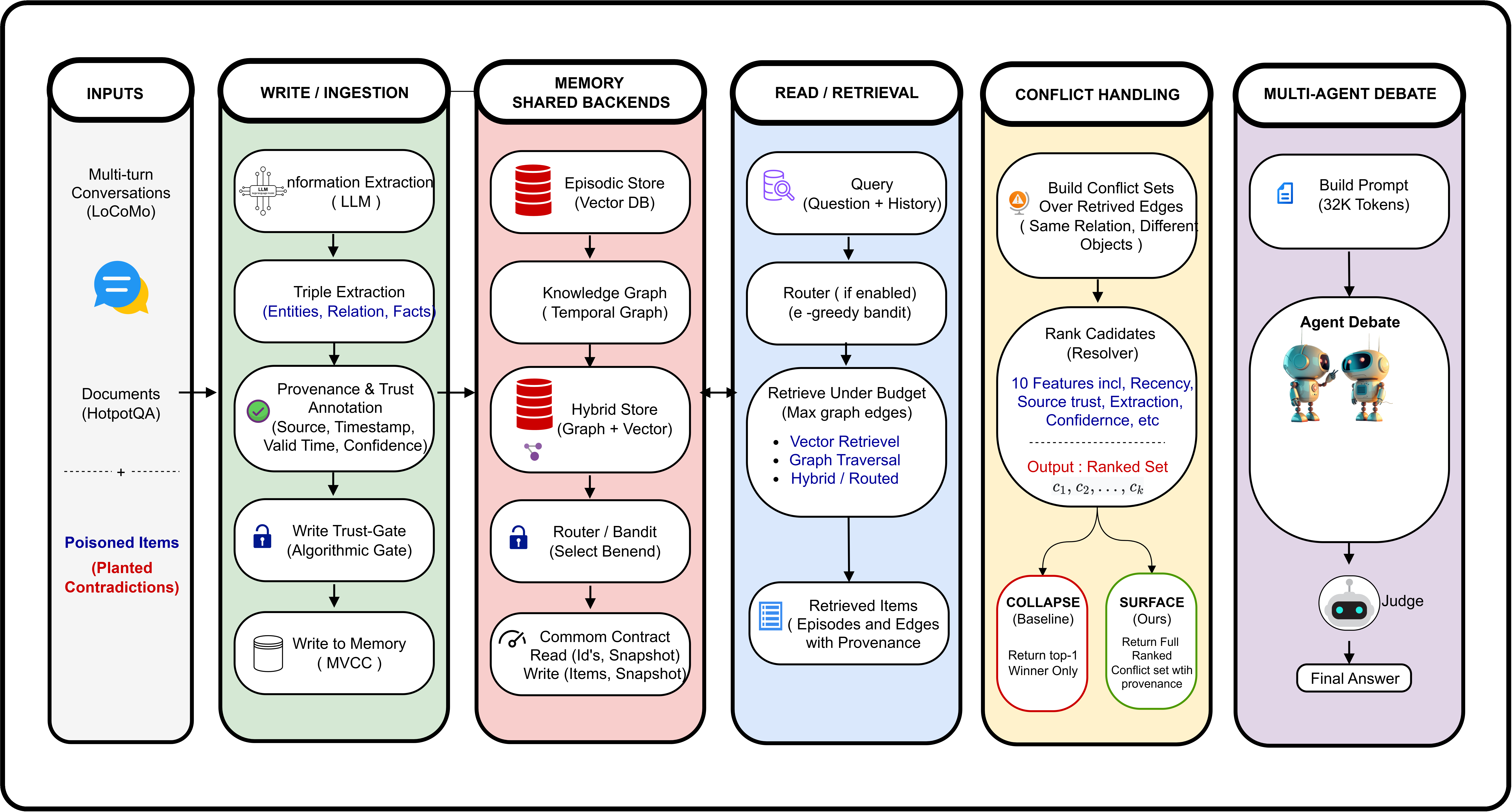}
\caption{DebateGraph, left to right. Conversations and documents are ingested with planted
contradictions mixed in; extraction produces provenance-annotated triples; the write
trust-gate admits, rejects or supersedes them into a bi-temporal store shared by all
backends. At query time a router retrieves under a hard token budget, and the resolver
groups retrieved edges that assert different objects for the same relation into ranked
conflict sets. \textbf{The one switch this paper is about sits at the end of that
stage}: \textsc{collapse} passes only the top-ranked claim, \textsc{surface} passes the
whole ranked set with its provenance. Everything downstream, the three agents, the two
rounds and the judge, is identical in both arms. Each evaluated mechanism is one
independent switch, so a single ablation changes exactly one thing.}
\label{fig:architecture}
\end{figure}

\Cref{fig:architecture} shows the system; we describe each mechanism and the switch that
removes it. Full parameterisation is in \cref{app:method}.

\textbf{Store and retrieval.} Each dialogue turn becomes one episode stamped with its
session date; triples are extracted by a batched LLM pass and cached, each edge retaining a
pointer to its episode so provenance is exact. Four retrieval modes are available:
\textsc{vector} (ANN over local embeddings~\citep{johnson2019faiss,xiao2024bge}),
\textsc{hop1}/\textsc{hopk} ($k$-hop traversal from seed entities), and \textsc{hybrid}.
Paths are scored by the product of edge trust and vector scores by $\sqrt{\rhotrust}$: this
is the \emph{write$\rightarrow$read coupling}, trust assigned on write driving ranking on
read. Switches: \texttt{router.mode}, \texttt{trust\_weighting}.

\textbf{Write trust-gate.} For a candidate edge we compute, with no LLM call,
$\rhotrust = c_{\text{writer}}\cdot(w_g g + w_b b + w_r \kappa)\cdot(1-\gamma)$, where $g$ is
endpoint groundedness, $b$ a depth-bounded BFS compatibility term, $\kappa$ relation
compatibility, $c$ a multiplicatively-updated writer credibility and $\gamma$ a contradiction
penalty. The edge commits iff $\rhotrust \ge \operatorname{median}$ of previously-committed
scores. Switch: \texttt{trust\_gate}, which governs admission \emph{and}
invalidate-on-supersede, a scope that turns out to matter (\cref{sec:gate}).

\textbf{Conflict resolution.} The resolver groups retrieved edges by $(h,r)$ and ranks
competing objects lexicographically by (temporal validity, $\rhotrust$, cross-writer
support). Two filters keep it honest on noisy extractions: only single-valued relations may
be contradicted (functionality computed over corpus edges alone), and objects asserted at the
same time are elaborations, not disagreements. Under \textbf{surface} the whole ranked set
enters the prompt with writer, trust, evidence count and supersession status; under
\textbf{collapse} only $\pi$'s top claim does, and losing claims are suppressed. Switches:
\texttt{conflict.surface}, \texttt{conflict.temporal\_validity}.

\textbf{Consistency.} Round $r$ reads a view $\mathcal{V}_r$ pinned at the round boundary;
concurrent agent writes buffer in staging and merge at the next boundary. Switch:
\texttt{mvcc}.

\section{Experimental Setup}
\label{sec:setup}

\textbf{Corpora.} LoCoMo~\citep{maharana2024locomo}: 10 conversations, $5{,}878$ episodes,
$9{,}769$ edges, giving \textsc{multihop-temporal} (categories 1--2) and \textsc{single-hop}
(category 4), $n{=}100$ stratified items each. For external validity we add
HotpotQA-distractor~\citep{yang2018hotpotqa}: one question's 10-paragraph set is one memory
partition, one sentence is one episode, and the source article title is the writer. Wikipedia
carries no event time, so $t_{\text{valid}}$ there is a \emph{synthetic ingestion clock};
that corpus therefore validates a reduced mechanism, with provenance intact and
bi-temporality synthetic (\cref{app:hotpot}).

\textbf{DebateMem-Conflict.} For an item whose gold answer is locatable as a single-valued
edge object, we synthesise a contradicting edge on the same $(h,r)$ with a
\emph{type-consistent} object drawn from the same relation elsewhere, attributed to a fourth
non-debating writer and stamped $30$ days later, so recency alone favours the lie. $122$ of
$603$ items ($20.2\%$) qualify. \texttt{source\_type} is \emph{never} rendered into a prompt:
labelling an injected memory as injected would hand agents the answer.

\textbf{Retrieval cap.} $k$-hop expansion truncates to $\texttt{graph\_max\_edges}$ candidate
edges ordered by hop distance then path trust. This was declared as $40$ before the runs and
is used everywhere except \cref{tab:regimes} and its Phase 1 control, which use $200$. The cap
governs which edges are \emph{candidates}; the $1{,}500$-token budget independently governs
what reaches the prompt. \Cref{sec:conflict} shows why the distinction matters.

\textbf{Models and protocol.} \texttt{api.txt} exposes two chat models and no embedding
endpoint; we use \texttt{gpt-oss-120b}~\citep{agarwal2025gptoss} throughout because the
70B alternative sustains $0.02$ req/s here (\cref{app:models}). $N{=}3$ agents, $R{=}2$
rounds, one judge $=7$ calls per item; budget $B{=}1500$ tokens ($32$K for full-context, which
then omits nothing). Sampling is $T{=}0.2$, $p{=}0.7$.

\textbf{Statistics.} All contrasts share item sets and are paired: $10{,}000$-resample
bootstrap~\citep{efron1987bootstrap} with two-sided Wilcoxon, and exact McNemar on
attack-success discordance. An effect is \emph{established} only when the $95\%$ bootstrap CI
excludes zero. We pre-specify two subgroups: the \emph{poisoned} items, since a treatment
firing only on contradictions cannot move items without one; and the \emph{discriminating}
items, where a conflict actually reached the context, since otherwise both arms receive
identical prompts and are tied by construction. The second filter uses the retriever alone,
no outcome information, and is applied identically to every corpus.

Those ties also make the Wilcoxon rule load-bearing, so we state it rather than inherit it.
At the pre-declared cap $94$ of $100$ pairs are exactly zero, and the two standard
treatments disagree: keeping them and splitting their ranks (\texttt{zsplit}) gives
$p{=}0.2595$ on the Phase 3 contrast, discarding them gives $p{=}0.0260$ on the same data.
We report both and rest no claim on either, because a pair that \emph{could not} have
differed is not evidence that it did not. The bootstrap CI on the pre-specified subgroup is
the test we declared, and it is what the asterisks in our tables mark.

\section{Results}
\label{sec:results}

\begin{table}[t]
\centering\small
\caption{Pre-declared gates, recomputed from stored summaries by \texttt{scripts/gates.py}.
Four of seven do not pass. $^\dagger$ The threshold is missed decisively (graph \emph{loses}
$0.150$ F1), so the verdict is \refuted, not \pass; the plan pre-declared that refuting
KG-as-default redirects the contribution to the routing policy, so the protocol advanced on a
negative result. $^\ddagger$ Accuracy preserved, but not load-bearing (\cref{app:mvcc}).}
\label{tab:gates}
\begin{tabular}{clp{6.6cm}c}
\toprule
Phase & Mechanism & Pre-declared threshold & Verdict \\
\midrule
0 & harness \& ceiling & both slices run, all metrics logged & \pass \\
1 & KG as default      & graph adds $\ge3$ F1 pts at $\le2\times$ tokens & \refuted$^\dagger$ \\
2 & write trust-gate   & cuts ASR at $<5\%$ latency, ${\approx}0$ tokens & \incon \\
3 & conflict surfacing & beats collapse on final-answer accuracy & \pass \\
4 & retrieval router   & beats \emph{both} fixed arms on acc/1K & \fail \\
5 & snapshot isolation & serial-equivalent accuracy preserved & \pass$^\ddagger$ \\
6 & ablation grid      & every ablation moves the number it isolates & \fail \\
\bottomrule
\end{tabular}
\end{table}

\subsection{Phase 0--1: the ceiling, and whether structure pays}
\label{sec:phase1}

Full context recovers essentially all gold evidence at a $32$K budget and omits nothing, so
it is a genuine accuracy \emph{and} cost ceiling: F1 $0.503$ on \textsc{multihop-temporal} and
$0.662$ on \textsc{single-hop}, at ${\sim}147$K tokens/query. Against it the graph backend
does not merely fail to add its required $3$ points: it loses $0.150$ F1 (CI
$[-0.216,-0.085]$, $p<10^{-4}$) and $0.240$ on \textsc{single-hop}, while spending
$1.75\times$ the tokens of vector retrieval ($0.304$ F1 at $6{,}405$ tokens). It is
simultaneously budget-saturated and evidence-poor ($0.30$ gold-evidence recall against
vector's $0.50$).

Because the retrieval cap turns out to be decisive for the conflict mechanism
(\cref{sec:conflict}), we re-measured this row at a cap of $200$ to test whether the graph
deficit is itself a truncation artefact. It is not, and the result sharpens the diagnosis:
recall rises $0.30 \to \mathbf{0.76}$ while F1 moves $0.154 \to 0.155$ and tokens/query
$11{,}199 \to 11{,}413$. Retrieving two and a half times as much of the right evidence buys
one thousandth of an F1 point. The graph backend's problem is not coverage but the form in
which evidence is delivered: a budget-sized block of triples is worse debate context than the
same budget of verbatim dialogue, even when the triples demonstrably contain the answer.
Upstream, only $20.2\%$ of these questions have an answer expressible as a single-valued edge
object. Per the pre-declared gate, this redirects the contribution to the routing policy.

\subsection{Phase 2: the write trust-gate does not work, and why}
\label{sec:gate}

Implemented as specified the gate rejects $0/122$ poisoned writes. We made three principled
repairs and measured each (\cref{app:gate}); together they move rejection to $3/122$. The
residual cause is structural. A type-consistent distractor is by construction composed of
objects the corpus already mentions, so \textbf{$100\%$ of poison objects are entities already
in the graph, against $53\%$ of gold objects}. Endpoint groundedness therefore scores the lie
\emph{above} a truthful first mention, and below only a restatement of an existing triple. The
information that separates them is \emph{relative} (candidate versus incumbent on the same
key), which is exactly what a global scalar threshold discards and what the Phase 3 resolver
consumes.

At the pre-declared cap the gate is free ($+2.7\%$ latency, $-15$ tokens/query) and inert: it
rejects $1$ of $51$ attacks, changes F1 by $-0.019$ (CI $[-0.047,+0.002]$), and changes ASR by
\emph{exactly} $0.000$. We did not weaken the attack to rescue it; sampling distractors that
introduce unseen entities would make the gate look effective by making the attack unrealistic.

\textbf{With contradictions reachable, the gate is not inert but harmful.} Re-measured at a cap
of $200$ with every poisonable item attacked (\cref{tab:regimes}), enabling the gate costs
$0.075$ F1 (CI $[-0.133,-0.018]$, Wilcoxon $p{=}0.039$) and moves attacks the wrong way,
$7 \to 9$ of $122$. The run's own write log identifies the mechanism: \texttt{trust\_gate}
governs invalidate-on-supersede as well as admission, and over these $122$ items it fires
\textbf{$123$ invalidations against $3$ rejections}. Almost none of its effect is the admission
decision it was built for; almost all is supersession. A gate that cannot separate poison from
truth still \emph{acts}, and when contradictions reach retrieval it spends that heuristic on
live claims and invalidates true ones.

\subsection{Phase 3: surfacing conflict beats collapsing it}
\label{sec:conflict}

This is the one mechanism that survives. On the graph backend, surfacing gains $+0.043$ F1
overall (CI $[+0.010,+0.084]$) and $+0.084$ on the poisoned subgroup (CI $[+0.022,+0.160]$);
on the router backend, $+0.077$ and $+0.195$. Two independent substrates, same direction, both
CIs excluding zero. Attacks fall $5\to3$ of $51$, but discordant pairs are few and McNemar is
not significant, so at this configuration we claimed an accuracy effect and no established ASR
reduction.

\textbf{That measurement runs through a retrieval bottleneck.} Conflict sets are built only
over edges retrieval returned, and $k$-hop truncates to $40$ ordered by hop distance then path
trust, never by relevance to the question. On a hub entity such as a speaker's own name, the
edge carrying the gold answer routinely misses the cut. Sweeping the cap offline, with no LLM
calls, the number of poisoned items whose planted contradiction is ever retrieved rises
$12/51 \to 27/51 \to 40/51$ at caps of $40, 100, 200$. At the pre-declared setting the
mechanism cannot act on three quarters of the attacks it was built for, and those items are
tied by construction rather than by observation.

\begin{table}[t]
\centering\small
\caption{Every conflict-dependent contrast, at the pre-declared cap of $40$ and again at
$200$, same paired procedure. Positive means the mechanism helps. Only the two surfacing rows
are mechanisms we claim; the rest are the paper's null results, and two stop being null.
$^\ast$ denotes a $95\%$ bootstrap CI excluding zero.}
\label{tab:regimes}
\begin{tabular}{lrrl}
\toprule
Contrast ($\Delta$F1, mechanism on $-$ off) & cap $40$ & cap $200$ & at cap $200$ \\
\midrule
surfacing, graph backend        & $+0.043^\ast$ & \best{$+0.340^\ast$} & helps \\
surfacing, router backend       & $+0.077^\ast$ & \best{$+0.143^\ast$} & helps \\
\midrule
write trust-gate                & $-0.019$ & $-0.075^\ast$ & \textbf{harms} \\
router provenance features      & $-0.002$ & $-0.096^\ast$ & \textbf{harms} \\
\midrule
trust-weighted retrieval        & $-0.059$ & $-0.065$ & inert \\
temporal validity in ranking    & $+0.001$ & $-0.004$ & inert \\
write gate (full-system ablation) & $+0.012$ & $+0.004$ & inert \\
\bottomrule
\end{tabular}
\end{table}

\textbf{With the contradiction reachable, both halves of the claim are established.} Re-running
the same contrast at a cap of $200$, poisoning every poisonable item, three things move
together and interpret each other. Cost does not: tokens/query goes $11{,}182 \to 11{,}318$
(collapse) and $11{,}224 \to 11{,}550$ (surface), because the cap governs candidates and the
budget still governs the prompt. The \emph{collapse} arm does not move either,
$0.1165 \to 0.1158$, so nothing about the wider net helps a debate shown one claim. And the
attack becomes markedly more effective once retrievable, ASR $0.098 \to 0.180$, the honest
reading of what the cap was hiding: not a weak attack, an unreachable one.

Against that stronger attack, surfacing raises F1 from $0.116$ to $\mathbf{0.456}$
($\Delta{=}+0.340$, CI $[+0.267,+0.414]$, Wilcoxon $p<10^{-4}$) and cuts attacks from $22$ to
$7$ of $122$, with $16$ discordant pairs favouring surface against $1$ favouring collapse
(\textbf{McNemar exact $p{=}0.0003$}). Discriminating items rise from $13/100$ to $109/122$.
We therefore claim both halves (a gain in accuracy and a reduction in attack success) as
established, and report the pre-declared cap-$40$ numbers unchanged rather than replacing them.

\textbf{The regime change does not spare the null results.} Raising the cap is not a knob that
only flatters the mechanism we claim, and \cref{tab:regimes} is the check. Both surfacing
contrasts grow, but two mechanisms that read as no-ops at cap $40$ are actively harmful once
contradictions reach the prompt: the write trust-gate ($-0.075$, $p{=}0.039$) and the router's
provenance features ($-0.096$, CI $[-0.149,-0.043]$, $p{=}0.030$). Three remain inert in both
regimes. \textbf{This is the honest cost of the correction: the same measurement that
multiplied our positive result by eight also converted two null results into negative ones,
and we would have had to report the second even if only the first had been convenient.}

\subsection{Cross-corpus replication}
\label{sec:xcorpus}

To test whether the surviving mechanism depends on memory being conversational, we replicated
the Phase 3 contrast on HotpotQA-distractor. The direction replicates: on the discriminating
subset, $+0.111$ (CI $[+0.010,+0.232]$) against LoCoMo's $+0.329$, with $33/100$ discriminating
items against $13/100$. On the full sample the HotpotQA interval contains zero. \textbf{We claim
a directional replication of the accuracy effect on a second corpus, established on the
pre-specified discriminating subset and not on the full sample.} The ASR endpoint replicates in
direction ($3\to2$ of $67$) and establishes nothing.

We chose this corpus for a structural reason. A HotpotQA partition is one question's ten
paragraphs ($67$--$176$ edges against LoCoMo's ${\sim}729$), so the retrieval cap cannot bind,
and conflict reachability is identical at caps of $40$ and $200$ ($34/67$ either way). A
replication on a corpus with the same bottleneck would have inherited the confound rather than
tested past it. Two properties of the port cut against the effect: $67\%$ of items are
poisonable (against $20.2\%$), but the more diverse relation vocabulary makes same-relation
distractors scarcer, so some injected lies are type-inconsistent and \emph{easier} to reject in
either arm, biasing the contrast toward zero. Ceilings also differ (HotpotQA's collapse arm
already answers at $0.388$ against LoCoMo's $0.116$), so the mechanism has less room.

\subsection{Router, ablations and baselines}
\label{sec:ablation}

The bandit significantly beats the heuristic it was meant to replace ($+0.063$, CI
$[+0.023,+0.111]$), which was the pre-declared ship decision, but not always-vector ($+0.010$,
CI $[-0.013,+0.040]$, marginally lower acc/1K). The gate fails. The interesting part is
\emph{how}: the bandit converges to selecting vector on $93\%$ of queries, rediscovering by
online search the conclusion Phase 1 gave directly, and the $\lambda$ sweep is flat because the
arms do not trade off (\cref{app:router}). Snapshot isolation preserves serial-equivalent
accuracy but is not load-bearing at this debate topology (\cref{app:mvcc}). The memory service
also runs unmodified inside AutoGen~\citep{wu2023autogen} and reproduces the in-house ordering
exactly, which we report as transfer of the abstraction and explicitly not as a gain
(\cref{app:autogen}).

\section{Discussion}

\begin{figure}[t]
\centering
\includegraphics[width=\textwidth]{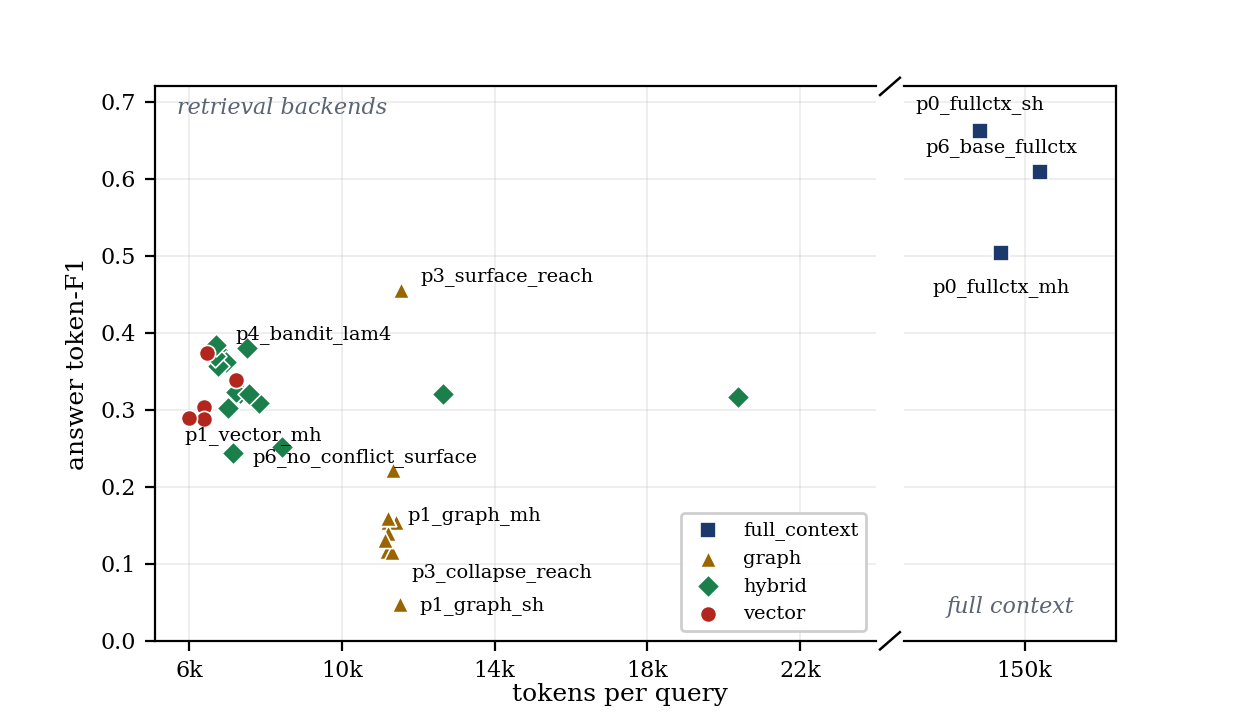}
\caption{Accuracy--cost frontier over the 37 LoCoMo runs (the two HotpotQA runs are excluded
as a different corpus). The $x$-axis is broken: the two regimes are an order of magnitude apart
in cost, so each is drawn at its own scale and the empty middle is the point rather than an
artefact. No configuration is simultaneously cheap and accurate, and the $\lambda$ sweep
collapses into a single cluster rather than tracing a trade-off. The exception is
\texttt{p3\_surface\_reach} at F1 $0.456$ for $11.5$K tokens (\cref{tab:regimes}), which lifts
the retrieval ceiling by $0.07$ over every other retrieval run at no extra cost. Only runs
discussed in the text are labelled.}
\label{fig:pareto}
\end{figure}

\textbf{Why not just use full context?} On this corpus that is the right baseline and we do not
beat it. The largest LoCoMo conversation is $23{,}864$ tokens, so every partition fits inside
the $32$K budget, full context omits nothing, and it wins by $0.05$--$0.21$ F1 over the best
retrieval configuration. We state this plainly because the alternative (reporting retrieval
numbers and letting the reader assume the window was the constraint) is the more common
failure. Two things follow. The honest motivation for retrieval here is cost, not capability:
$147$K tokens/query against $11$K, a $13\times$ premium for $10\%$ more F1
(\cref{fig:pareto}). And the conflict
result does not depend on the window at all: collapse and surface see the \emph{same} retrieved
set at the \emph{same} budget, differing only in whether the contradiction inside it survives. A
system with an infinite window and a store that reconciles on write would still hand its agents
the reconciled answer, and would still be wrong in the way \cref{tab:regimes} measures.

\textbf{The storage layer should not adjudicate.} Our positive result and our sharpest negative
one are the same finding twice. Surfacing conflict helps; and when the resolver ranks badly (as
it does under recency-first ordering) final accuracy is nonetheless unharmed, because the agents
override it. Meanwhile the one component that \emph{does} adjudicate on write, the trust-gate,
costs accuracy once it has live claims to act on. A debate system already contains machinery for
adjudicating competing claims. Asking the storage layer to do that job with a heuristic tie-break
is both unnecessary and, under an adversary who controls timestamps, exploitable.
\emph{Preserve the disagreement; let the reasoner reason.}

\textbf{Groundedness is the wrong prior for write admission.} Plausibility scores built on ``does
this connect to what I already know'' are, against an adversary who composes lies from known
entities, better satisfied by lies than by truths. Any admission policy of this family inherits
the flaw; admission control for agent memory likely needs corroboration or source-level signals,
not topology.

\section{Limitations}
\label{sec:limitations}

\textbf{Not every run shares the reachable regime.} \Cref{tab:regimes} re-measures all seven
conflict-dependent contrasts at cap $200$, and \cref{sec:phase1} re-measures the Phase 1
deficit, so every mechanism claim exists in both regimes. What stays at cap $40$ are runs where
the cap cannot be load-bearing because no contradiction is at stake: the $\lambda$ sweep, the
agent-count and adversary-rate sweeps, MVCC, the full-context baselines and the AutoGen
transfer. We judge those unaffected rather than having verified it, and that judgement is the
residual form of this limitation.

\textbf{Statistical power on the weaker conditions.} The ASR reduction is established at cap
$200$ ($22\to7$ of $122$, $p{=}0.0003$) but not at cap $40$ ($5\to3$ of $51$, $p{=}0.63$) nor on
HotpotQA ($3\to2$ of $67$, $p{=}1.00$), where too few attacks land for any test to resolve. ASR
statements outside \cref{tab:regimes} remain directional.

\textbf{Single model family.} All results use \texttt{gpt-oss-120b}, so we cannot separate
model-specific from mechanism-level effects. \Cref{sec:xcorpus} removes the single-corpus half of
this limitation, but the model axis is untested: we attempted the replication and abandoned it on
measurement, since under this paper's load a single debate item did not complete in $35$ minutes
on the 70B endpoint (\cref{app:models}).

\textbf{Memory never exceeds the context window.} Every LoCoMo partition fits in $32$K tokens, so
full context is always available here and retrieval is motivated by cost rather than necessity.
The conflict result is independent of this (both arms share one retrieved set and one
budget), but we do not exhibit the regime where memory genuinely overflows, which is where a
storage layer's decision to reconcile becomes unavoidable rather than merely cheap.

\textbf{Benchmark scope.} DebateMem-Conflict covers the $20.2\%$ of items whose answers are
single-valued edge objects, biasing toward entity-like factual questions and away from temporal
ones. Injections are template-generated for determinism, so they test contradiction handling
rather than fluent adversarial paraphrase. A single judge sits between memory and score.

\section{Conclusion}

We built a debate-memory service in which five mechanisms from the agent-memory literature could
be switched off one at a time behind one contract, declared a threshold for each before running,
and measured all of them on 39 controlled runs. One survived: preserving contradictions and
handing the ranked conflict set to the debate improves F1 by $+0.340$ and cuts attack success
from $22$ to $7$ of $122$ poisoned items at equal token cost, replicating on two retrieval
backends and directionally on a second corpus. Four did not, and two did worse than nothing: with
contradictions reachable, the write trust-gate and the router's provenance features each cost
accuracy rather than merely failing to add it.

The largest correction in this paper is one we made to ourselves. Our pre-declared retrieval cap
left three quarters of planted contradictions unretrievable, and the surviving mechanism measured
at an eighth of its size with its robustness half indistinguishable from noise. The
instrumentation that found this (per-item conflict counts, logged whether or not anything used
them) cost nothing and was worth more than any single experiment we ran. A mechanism can only be
measured where it is able to act, and checking that it \emph{can} act is a separate obligation
from checking whether it helps.

\label{endofcontent}   
\bibliographystyle{plainnat}
\bibliography{references}

\begin{thebibliography}{27}
\providecommand{\natexlab}[1]{#1}
\providecommand{\url}[1]{\texttt{#1}}
\expandafter\ifx\csname urlstyle\endcsname\relax
  \providecommand{\doi}[1]{doi: #1}\else
  \providecommand{\doi}{doi: \begingroup \urlstyle{rm}\Url}\fi

\bibitem[Bernstein et~al.(1987)Bernstein, Hadzilacos, and
  Goodman]{bernstein1987concurrency}
Philip~A. Bernstein, Vassos Hadzilacos, and Nathan Goodman.
\newblock \emph{Concurrency Control and Recovery in Database Systems}.
\newblock Addison-Wesley, 1987.

\bibitem[Cattan et~al.(2025)]{cattan2025dragged}
Arie Cattan et~al.
\newblock Dragged into conflicts: Detecting and addressing conflicting sources
  in search-augmented llms.
\newblock \emph{arXiv preprint arXiv:2506.08500}, 2025.

\bibitem[Chen et~al.(2024)Chen, Xiang, Xiao, Song, and Li]{chen2024agentpoison}
Zhaorun Chen, Zhen Xiang, Chaowei Xiao, Dawn Song, and Bo~Li.
\newblock {AgentPoison}: Red-teaming {LLM} agents via poisoning memory or
  knowledge bases.
\newblock In \emph{Advances in Neural Information Processing Systems
  (NeurIPS)}, 2024.

\bibitem[Chhikara et~al.(2025)Chhikara, Khant, Aryan, Singh, and
  Yadav]{chhikara2025mem0}
Prateek Chhikara, Dev Khant, Saket Aryan, Taranjeet Singh, and Deshraj Yadav.
\newblock Mem0: Building production-ready {AI} agents with scalable long-term
  memory.
\newblock \emph{arXiv preprint arXiv:2504.19413}, 2025.

\bibitem[Dong et~al.(2014)Dong, Gabrilovich, Heitz, Horn, Lao, Murphy,
  Strohmann, Sun, and Zhang]{dong2014knowledgevault}
Xin Dong, Evgeniy Gabrilovich, Geremy Heitz, Wilko Horn, Ni~Lao, Kevin Murphy,
  Thomas Strohmann, Shaohua Sun, and Wei Zhang.
\newblock Knowledge vault: A web-scale approach to probabilistic knowledge
  fusion.
\newblock In \emph{Proceedings of the 20th ACM SIGKDD International Conference
  on Knowledge Discovery and Data Mining (KDD)}, 2014.

\bibitem[Du et~al.(2024)Du, Li, Torralba, Tenenbaum, and
  Mordatch]{du2024debate}
Yilun Du, Shuang Li, Antonio Torralba, Joshua~B. Tenenbaum, and Igor Mordatch.
\newblock Improving factuality and reasoning in language models through
  multiagent debate.
\newblock In \emph{Proceedings of the 41st International Conference on Machine
  Learning (ICML)}, 2024.

\bibitem[Edge et~al.(2024)Edge, Trinh, Cheng, Bradley, Chao, Mody, Truitt, and
  Larson]{edge2024graphrag}
Darren Edge, Ha~Trinh, Newman Cheng, Joshua Bradley, Alex Chao, Apurva Mody,
  Steven Truitt, and Jonathan Larson.
\newblock From local to global: A graph {RAG} approach to query-focused
  summarization.
\newblock \emph{arXiv preprint arXiv:2404.16130}, 2024.

\bibitem[Efron(1987)]{efron1987bootstrap}
Bradley Efron.
\newblock Better bootstrap confidence intervals.
\newblock \emph{Journal of the American Statistical Association}, 82:\penalty0
  171--185, 1987.

\bibitem[Gal{\'a}rraga et~al.(2013)Gal{\'a}rraga, Teflioudi, Hose, and
  Suchanek]{galarraga2013amie}
Luis~Antonio Gal{\'a}rraga, Christina Teflioudi, Katja Hose, and Fabian~M.
  Suchanek.
\newblock {AMIE}: Association rule mining under incomplete evidence in
  ontological knowledge bases.
\newblock In \emph{Proceedings of the 22nd International Conference on World
  Wide Web (WWW)}, 2013.

\bibitem[Guti{\'e}rrez et~al.(2024)Guti{\'e}rrez, Shu, Gu, Yasunaga, and
  Su]{gutierrez2024hipporag}
Bernal~Jim{\'e}nez Guti{\'e}rrez, Yiheng Shu, Yu~Gu, Michihiro Yasunaga, and
  Yu~Su.
\newblock {HippoRAG}: Neurobiologically inspired long-term memory for large
  language models.
\newblock In \emph{Advances in Neural Information Processing Systems
  (NeurIPS)}, 2024.

\bibitem[Johnson et~al.(2019)Johnson, Douze, and J{\'e}gou]{johnson2019faiss}
Jeff Johnson, Matthijs Douze, and Herv{\'e} J{\'e}gou.
\newblock Billion-scale similarity search with {GPU}s.
\newblock \emph{IEEE Transactions on Big Data}, 7\penalty0 (3), 2019.

\bibitem[Li et~al.(2010)Li, Chu, Langford, and Schapire]{li2010linucb}
Lihong Li, Wei Chu, John Langford, and Robert~E. Schapire.
\newblock A contextual-bandit approach to personalized news article
  recommendation.
\newblock In \emph{Proceedings of the 19th International Conference on World
  Wide Web (WWW)}, 2010.

\bibitem[Liang et~al.(2024)Liang, He, Jiao, Wang, Wang, Wang, Yang, Shi, and
  Tu]{liang2024divergent}
Tian Liang, Zhiwei He, Wenxiang Jiao, Xing Wang, Yan Wang, Rui Wang, Yujiu
  Yang, Shuming Shi, and Zhaopeng Tu.
\newblock Encouraging divergent thinking in large language models through
  multi-agent debate.
\newblock In \emph{Proceedings of the 2024 Conference on Empirical Methods in
  Natural Language Processing (EMNLP)}, 2024.

\bibitem[Liu et~al.(2023)Liu, Zhang, Li, Liu, and Yang]{liu2023dylan}
Zijun Liu, Yanzhe Zhang, Peng Li, Yang Liu, and Diyi Yang.
\newblock Dynamic {LLM}-agent network: An {LLM}-agent collaboration framework
  with agent team optimization.
\newblock \emph{arXiv preprint arXiv:2310.02170}, 2023.

\bibitem[Maharana et~al.(2024)Maharana, Lee, Tulyakov, Bansal, Barbieri, and
  Fang]{maharana2024locomo}
Adyasha Maharana, Dong-Ho Lee, Sergey Tulyakov, Mohit Bansal, Francesco
  Barbieri, and Yuwei Fang.
\newblock Evaluating very long-term conversational memory of {LLM} agents.
\newblock In \emph{Proceedings of the 62nd Annual Meeting of the Association
  for Computational Linguistics (ACL)}, 2024.

\bibitem[Meng et~al.(2026)]{meng2026equimem}
Yuqiao Meng et~al.
\newblock Equimem: Calibrating shared memory in multi-agent debate via
  game-theoretic equilibrium.
\newblock \emph{arXiv preprint arXiv:2605.09278}, 2026.

\bibitem[{OpenAI}(2025)]{agarwal2025gptoss}
{OpenAI}.
\newblock gpt-oss-120b \& gpt-oss-20b model card.
\newblock \emph{arXiv preprint arXiv:2508.10925}, 2025.

\bibitem[Packer et~al.(2023)Packer, Wooders, Lin, Fang, Patil, Stoica, and
  Gonzalez]{packer2023memgpt}
Charles Packer, Sarah Wooders, Kevin Lin, Vivian Fang, Shishir~G. Patil, Ion
  Stoica, and Joseph~E. Gonzalez.
\newblock {MemGPT}: Towards {LLM}s as operating systems.
\newblock \emph{arXiv preprint arXiv:2310.08560}, 2023.

\bibitem[Rasmussen et~al.(2025)Rasmussen, Paliychuk, Beauvais, Ryan, and
  Chalef]{rasmussen2025zep}
Preston Rasmussen, Pavlo Paliychuk, Travis Beauvais, Jack Ryan, and Daniel
  Chalef.
\newblock Zep: A temporal knowledge graph architecture for agent memory.
\newblock \emph{arXiv preprint arXiv:2501.13956}, 2025.

\bibitem[Tian et~al.(2026)Tian, Feng, Zhao, Zhu, Yan, and Han]{tian2026madmm}
Hongduan Tian, Xiao Feng, Ziyuan Zhao, Xiangyu Zhu, Rolan Yan, and Bo~Han.
\newblock Multi-agent debate with memory masking.
\newblock In \emph{International Conference on Learning Representations
  (ICLR)}, 2026.

\bibitem[Wang et~al.(2025)Wang, Prasad, Stengel-Eskin, and
  Bansal]{wang2025madamrag}
Han Wang, Archiki Prasad, Elias Stengel-Eskin, and Mohit Bansal.
\newblock Retrieval-augmented generation with conflicting evidence.
\newblock In \emph{Conference on Language Modeling (COLM)}, 2025.

\bibitem[Wang et~al.(2023)Wang, Wei, Schuurmans, Le, Chi, Narang, Chowdhery,
  and Zhou]{wang2023selfconsistency}
Xuezhi Wang, Jason Wei, Dale Schuurmans, Quoc~V. Le, Ed~H. Chi, Sharan Narang,
  Aakanksha Chowdhery, and Denny Zhou.
\newblock Self-consistency improves chain of thought reasoning in language
  models.
\newblock In \emph{International Conference on Learning Representations
  (ICLR)}, 2023.

\bibitem[Wu et~al.(2023)Wu, Bansal, Zhang, Wu, Li, Zhu, Jiang, Zhang, Zhang,
  Liu, Awadallah, White, Burger, and Wang]{wu2023autogen}
Qingyun Wu, Gagan Bansal, Jieyu Zhang, Yiran Wu, Beibin Li, Erkang Zhu,
  Li~Jiang, Xiaoyun Zhang, Shaokun Zhang, Jiale Liu, Ahmed~Hassan Awadallah,
  Ryen~W. White, Doug Burger, and Chi Wang.
\newblock Autogen: Enabling next-gen llm applications via multi-agent
  conversation.
\newblock \emph{arXiv preprint arXiv:2308.08155}, 2023.

\bibitem[Xiao et~al.(2023)Xiao, Liu, Zhang, Muennighoff, Lian, and
  Nie]{xiao2024bge}
Shitao Xiao, Zheng Liu, Peitian Zhang, Niklas Muennighoff, Defu Lian, and
  Jian-Yun Nie.
\newblock {C-Pack}: Packed resources for general chinese embeddings.
\newblock \emph{arXiv preprint arXiv:2309.07597}, 2023.

\bibitem[Yang et~al.(2018)Yang, Qi, Zhang, Bengio, Cohen, Salakhutdinov, and
  Manning]{yang2018hotpotqa}
Zhilin Yang, Peng Qi, Saizheng Zhang, Yoshua Bengio, William~W. Cohen, Ruslan
  Salakhutdinov, and Christopher~D. Manning.
\newblock {HotpotQA}: A dataset for diverse, explainable multi-hop question
  answering.
\newblock In \emph{Proceedings of the 2018 Conference on Empirical Methods in
  Natural Language Processing (EMNLP)}, 2018.

\bibitem[Zhang et~al.(2025)Zhang, Fu, Wan, Yu, Wang, and Ma]{zhang2025gmemory}
Guibin Zhang, Muxin Fu, Guancheng Wan, Miao Yu, Kun Wang, and Shuicheng Ma.
\newblock G-memory: Tracing hierarchical memory for multi-agent systems.
\newblock \emph{arXiv preprint arXiv:2506.07398}, 2025.

\bibitem[Zou et~al.(2024)Zou, Geng, Wang, and Jia]{zou2024poisonedrag}
Wei Zou, Runpeng Geng, Binghui Wang, and Jinyuan Jia.
\newblock {PoisonedRAG}: Knowledge corruption attacks to retrieval-augmented
  generation of large language models.
\newblock \emph{arXiv preprint arXiv:2402.07867}, 2024.

\end{thebibliography}

\appendix

\section{Method parameterisation}
\label{app:method}

\textbf{Trust score.} For a candidate edge,
\begin{equation}
\rhotrust = c_{\text{writer}} \cdot \underbrace{(w_g g + w_b b + w_r \kappa)}_{\text{base}}
            \cdot (1-\gamma), \qquad
b = \frac{1}{1 + d_{L_{\max}}(h,t)}, \qquad L_{\max} = \lceil \log|V| \rceil,
\label{eq:rho}
\end{equation}
where $g$ is endpoint groundedness (are $h$ and $t$ already known entities), $b$ a
depth-bounded BFS compatibility term, $\kappa$ relation compatibility, $c$ a
multiplicatively-updated writer credibility, and $\gamma$ a contradiction penalty. The edge
commits iff $\rhotrust \ge \operatorname{median}$ of previously-committed scores. $L_{\max}$ is
capped at 4 for latency. Corpus edges bypass the gate (transcript facts are ground truth by
construction), but their $\rhotrust$ is still computed, against the complete graph minus the
edge being scored, and seeds the running median.

\textbf{Conflict ranking.} The resolver ranks competing objects for a key $(h,r)$ by
\begin{equation}
\pi(\epsilon) = \big(\ \mathbb{1}[t_{\text{inval}} = \varnothing],\ t_{\text{valid}},\
\rhotrust,\ |\{\text{writers}\}|\ \big) \quad\text{lexicographically}.
\label{eq:rank}
\end{equation}
Two filters keep the mechanism honest on noisy extractions. \emph{Functionality}: only
single-valued relations can be contradicted, scored as $|\{h\}|/|\{(h,t)\}|$ over corpus edges
only, so a write cannot exempt itself by making its own relation look multi-valued.
\emph{Distinct time}: objects asserted in the same turn are elaborations, not disagreements;
near-duplicate objects (token-F1 $\ge 0.5$) are merged.

\textbf{Router.} Features $x(q)$ are structural complexity, temporal and comparative markers,
predicted hop count, seed-neighbourhood density, remaining budget, and a conflict-present flag.
We compare a fixed heuristic against disjoint LinUCB~\citep{li2010linucb} with reward
$\mathrm{rew} = u - \lambda\cdot\text{tokens}/\tau$.

\textbf{Budget.} Context is packed greedily to $B$ tokens with an explicit ``$n$ memories
omitted'' note appended; a mode never silently overflows. Budget tokens are counted with a
fixed local tokenizer so enforcement is deterministic and offline, while reported cost uses
server usage records.

\section{Model selection}
\label{app:models}

\begin{table}[htbp]
\centering\small
\caption{Endpoint measurement that determined model choice ($n{=}3$ sequential, $8$-way
concurrent, no errors; \texttt{results/model\_latency\_probe.json}). At $0.02$ req/s the
34-run matrix ($\sim$24k calls) would need ${>}300$ hours on the 70B endpoint. Absolute
latencies on this shared endpoint vary with load---an earlier probe recorded $39$\,s
sequential for the 70B model and $0.7$\,s for gpt-oss---but the two-order-of-magnitude
ratio was stable across every measurement we took.}
\label{tab:models}
\begin{tabular}{lrrr}
\toprule
model (both from \texttt{api.txt}) & seq.\ median & 8-way median & throughput \\
\midrule
\texttt{meta/llama-3.3-70b-instruct} & $228{,}050$\,ms & $172{,}575$\,ms & $0.02$ req/s \\
\texttt{openai/gpt-oss-120b}         & $1{,}539$\,ms   & $1{,}885$\,ms   & $3.63$ req/s \\
\bottomrule
\end{tabular}
\end{table}

\texttt{api.txt} for this environment exposes two chat models and no embedding endpoint, so
embeddings are a local GPU model~\citep{xiao2024bge}: this changes the embedder, not the
retrieval method. We re-probed the 70B endpoint when attempting the second-model-family
replication requested in \cref{sec:limitations}: it returned $0.035$ req/s on short prompts,
twice the rate recorded above, but under this paper's actual load ($1{,}500$-token contexts
and $900$-token generations) a single debate item did not complete in $35$ minutes, putting
even a $30$-item replication beyond $30$ hours. This is an environment limit rather than a
scientific choice.

\section{Phase 0--1: full backend table}
\label{app:phase1}

\begin{table}[htbp]
\centering\small
\caption{Backends measured alone. \textsc{multihop-temporal} and \textsc{single-hop},
$n{=}100$ each. ``omit'' is blocks dropped at the budget ceiling; ``ev.\ rec.'' is
gold-evidence recall in the assembled context.}
\label{tab:phase1}
\begin{tabular}{llrrrrrr}
\toprule
Slice & Backend & F1 & EM & tokens/q & acc/1K & ev.\ rec. & omit \\
\midrule
\multirow{3}{*}{\textsc{multihop-temp.}}
 & full-context      & \best{0.503} & 0.20 & $147{,}702$ & 0.0034 & \best{1.00} & 0.0 \\
 & vector            & 0.304 & 0.10 & $6{,}405$  & \best{0.0474} & 0.50 & 0.0 \\
 & graph ($k$-hop)   & 0.154 & 0.08 & $11{,}199$ & 0.0137 & 0.30 & 15.8 \\
\midrule
\multirow{3}{*}{\textsc{single-hop}}
 & full-context      & \best{0.662} & 0.39 & $145{,}567$ & 0.0045 & \best{1.00} & 0.0 \\
 & vector            & 0.288 & 0.17 & $6{,}399$  & \best{0.0449} & 0.29 & 0.0 \\
 & graph ($k$-hop)   & 0.048 & 0.02 & $11{,}533$ & 0.0041 & 0.06 & 16.0 \\
\bottomrule
\end{tabular}
\end{table}

\section{Gate calibration}
\label{app:gate}

Implemented exactly as specified, the gate rejects $0/122$ poisoned writes. We made three
principled repairs and measured each.

\emph{Scores were incomparable.} Corpus $\rhotrust$ was computed incrementally, each edge
scored against the partial graph at ingestion, so corpus scores (median $0.20$) sat far below
runtime candidate scores against the complete graph. Corpus edges are now scored against the
complete graph minus themselves: the view an unseen candidate gets. Threshold $0.200 \to
0.175$; rejection unchanged.

\emph{The threshold pooled incomparable populations.} A displacing write is now judged against
previously-committed \emph{displacing} writes rather than first mentions; rejection unchanged.

\emph{The contradiction penalty was flat.} Scaled by $1-\text{base}$, it vanishes for a
well-connected candidate. Made proportional to what is displaced,
$\rhotrust = c \cdot \text{base} \cdot (1 - \rhotrust_{\text{incumbent}})$, rejection moves
$0/122 \to 3/122$.

\begin{table}[htbp]
\centering\small
\caption{Offline gate calibration over all $122$ injections
(\texttt{results/gate\_calibration.json}). No formulation separates the attack. Poison and
truthful-restatement scores are well separated ($0.44$ vs.\ $0.83$); the threshold derived
from committed scores sits far below both.}
\label{tab:gate}
\begin{tabular}{llrrrr}
\toprule
$\gamma$ penalty & threshold rule & thresh. & poison $\rhotrust$ & truth $\rhotrust$ & poison rejected \\
\midrule
flat (as specified) & pooled median      & 0.200 & 0.583 & 0.825 & $0/122$ \\
flat                & conditioned median & 0.175 & 0.583 & 0.825 & $0/122$ \\
proportional        & pooled median      & 0.200 & 0.440 & 0.825 & $3/122$ \\
proportional        & conditioned median & 0.175 & 0.440 & 0.825 & $3/122$ \\
\bottomrule
\end{tabular}
\end{table}

\section{Phase 3 at the pre-declared cap}
\label{app:conflict40}

\begin{table}[htbp]
\centering\small
\caption{Collapse versus surface, holding everything else fixed. The poisoned subgroup is
pre-specified: the mechanism is a no-op on items with no contradiction. $^\ast$ denotes a
$95\%$ bootstrap CI excluding zero.}
\label{tab:conflict}
\begin{tabular}{llrrrr}
\toprule
Backend & Policy & F1 (all) & F1 (poisoned) & $\Delta$F1 poisoned & attacks \\
\midrule
\multirow{2}{*}{graph}
 & collapse & 0.117 & --- & \multirow{2}{*}{$+0.084^\ast$ \ci{+0.022}{+0.160}} & $5/51$ \\
 & surface  & \best{0.159} & --- & & $3/51$ \\
\midrule
\multirow{2}{*}{router (hybrid)}
 & collapse & 0.243 & --- & \multirow{2}{*}{$+0.195^\ast$ \ci{+0.108}{+0.292}} & $5/51$ \\
 & surface  & \best{0.321} & --- & & $3/51$ \\
\bottomrule
\end{tabular}
\end{table}

These are the cap-$40$ numbers, retained unchanged. \Cref{tab:regimes} in the body reports the
same contrasts with the contradiction reachable.

\section{HotpotQA port}
\label{app:hotpot}

One question's distractor set (2 gold $+$ 8 distractor paragraphs) is one memory partition, one
context sentence is one episode, and the source article title is the writer, so provenance
becomes ``which document asserted this''. \textsc{Comparison} questions are excluded: their
answers are yes/no and there is no type-consistent entity to inject against a boolean.
$5{,}918$ \textsc{bridge} questions are available; we sample $n{=}100$ with a fixed seed.

One mapping loses information. Wikipedia paragraphs carry no event time, so $t_{\text{valid}}$
is a \emph{synthetic ingestion clock}, one day per paragraph in presentation order. The
bi-temporal term therefore measures recency-of-ingestion, not recency-of-event, and the
replication validates a reduced form of the mechanism. Extraction, store construction, gate,
resolver, router and metrics are byte-for-byte the same code on both corpora, so any measured
difference is the corpus rather than the pipeline.

\begin{table}[htbp]
\centering\small
\caption{The Phase 3 contrast (surface $-$ collapse) on both corpora, same code and same
test procedure. \emph{Discriminating} items are those where retrieval actually returned a
conflict; on the rest the two arms build identical prompts and are tied by construction.
$^\ast$ denotes a $95\%$ bootstrap CI excluding zero.}
\label{tab:xcorpus}
\begin{tabular}{lrr}
\toprule
 & LoCoMo (dialogue) & HotpotQA (documents) \\
\midrule
poisonable items                 & $20.2\%$ & $67\%$ \\
discriminating items             & $13/100$ & $33/100$ \\
$\Delta$F1, all items            & $+0.043^\ast$ \ci{+0.010}{+0.082} & $+0.032$ \ci{-0.005}{+0.077} \\
$\Delta$F1, discriminating       & $+0.329^\ast$ \ci{+0.115}{+0.572} & $+0.111^\ast$ \ci{+0.010}{+0.232} \\
attacks                          & $5\to3$ of $51$ & $3\to2$ of $67$ \\
McNemar exact $p$                & $0.63$ & $1.00$ \\
\bottomrule
\end{tabular}
\end{table}

\section{Router arms, ablations and the accuracy--cost frontier}
\label{app:router}

\begin{table}[htbp]
\centering\small
\caption{Router arms on the conflict slice, $n{=}100$. No policy beats always-vector on
acc/1K, so the pre-declared gate fails. The bandit does significantly beat the heuristic.}
\label{tab:router}
\begin{tabular}{lrrrrl}
\toprule
Policy & F1 & tokens/q & acc/1K & ASR & mode selection \\
\midrule
always-vector      & 0.374 & $6{,}471$  & \best{0.0578} & 0.098 & --- \\
always-graph       & 0.131 & $11{,}145$ & 0.0117 & \best{0.039} & --- \\
heuristic router   & 0.321 & $7{,}584$  & 0.0423 & 0.059 & 73/20/0/7\,\% \\
bandit $\lambda{=}1$ & 0.377 & $6{,}738$ & 0.0560 & 0.078 & 93/2/3/2\,\% \\
bandit $\lambda{=}4$ & \best{0.384} & $6{,}718$ & 0.0572 & 0.078 & 93/2/3/2\,\% \\
\bottomrule
\end{tabular}\\[3pt]
\parbox{0.95\textwidth}{\footnotesize Mode selection is vector/1-hop/$k$-hop/hybrid. The
heuristic never selects $k$-hop; the bandit converges to vector on $93\%$ of queries.}
\end{table}

\begin{table}[htbp]
\centering\small
\caption{Single-mechanism ablations against the full system (router $+$ all mechanisms),
same 100 items. Only conflict surfacing moves answer F1 with a CI excluding zero; only
temporal validity moves conflict-resolution F1---in the wrong direction.}
\label{tab:main}
\begin{tabular}{lrrrrl}
\toprule
Configuration & F1 & $\Delta$F1 (95\% CI) & confl.\ F1 & ASR & ev.\ rec. \\
\midrule
full system                    & 0.321 & --- & 0.000 & 0.059 & 0.43 \\
\ \ $-$ conflict surfacing     & 0.243 & $-0.077^\ast$ \ci{-0.135}{-0.025} & 0.000 & 0.098 & 0.33 \\
\ \ $-$ temporal validity      & 0.320 & $-0.001$ \ci{-0.004}{+0.001} & \best{0.387} & 0.059 & 0.43 \\
\ \ $-$ trust weighting        & 0.380 & $+0.059$ \ci{-0.013}{+0.132} & 0.000 & 0.078 & 0.62 \\
\ \ $-$ provenance features    & 0.323 & $+0.002$ \ci{-0.000}{+0.006} & 0.000 & 0.059 & 0.43 \\
\ \ $-$ write trust-gate       & 0.309 & $-0.012$ \ci{-0.046}{+0.020} & 0.000 & 0.059 & 0.43 \\
\midrule
\multicolumn{6}{l}{\textit{Baselines (conflict slice)}}\\
full-context                   & \best{0.609} & $+0.225^\ast$ \ci{+0.144}{+0.308} & --- & \best{0.039} & 1.00 \\
vector-only (Mem0-style)       & 0.374 & --- & --- & 0.098 & 0.47 \\
graph-only (Zep/G-Memory-style)& 0.131 & --- & --- & 0.039 & 0.26 \\
trust-only                     & 0.289 & --- & --- & 0.078 & 0.35 \\
DebateGraph (bandit $\lambda{=}4$) & 0.384 & --- & --- & 0.078 & 0.50 \\
\bottomrule
\end{tabular}\\[3pt]
\parbox{0.95\textwidth}{\footnotesize $\Delta$F1 is relative to the full system except for
full-context, which is relative to DebateGraph. $^\ast$ CI excludes zero.}
\end{table}

The accuracy--cost frontier for these runs is \cref{fig:pareto} in the body.
The $\lambda$ sweep is flat (F1 $\in \{0.362, 0.357, 0.377, 0.370, 0.384\}$ for
$\lambda \in \{0.25, 0.5, 1, 2, 4\}$) with near-identical arm distributions: $\lambda$ can only
trace a Pareto frontier when the arms trade off, and here one arm dominates.

\section{Phase 5: round-level consistency}
\label{app:mvcc}

MVCC and a single lock-protected store produce serial-equivalent accuracy under concurrent
agent writes, so the pre-declared gate passes. The mechanism is nonetheless not load-bearing at
this debate topology: agents write reasoning episodes that assert no triples, so the staging
merge has little to reconcile. We report the gate as passed and the mechanism as inert, rather
than claiming a benefit the measurement does not show.

\section{Cross-framework transfer}
\label{app:autogen}

\begin{table}[htbp]
\centering\small
\caption{The same memory service under AutoGen~\citep{wu2023autogen}, unmodified
($n{=}60$). The behavioural signature transfers; the accuracy advantage does not exist in
either host.}
\label{tab:xframework}
\begin{tabular}{llrrrr}
\toprule
Host framework & Backend & F1 & tokens/q & acc/1K & ASR \\
\midrule
\multirow{2}{*}{AutoGen}  & vector & \best{0.339} & $7{,}226$ & \best{0.0469} & 0.143 \\
                          & full system & 0.251 & $8{,}436$ & 0.0298 & \best{0.114} \\
\midrule
\multirow{2}{*}{in-house} & vector & \best{0.374} & $6{,}471$ & \best{0.0578} & 0.098 \\
                          & full system & 0.321 & $7{,}584$ & 0.0423 & \best{0.059} \\
\bottomrule
\end{tabular}
\end{table}

The service runs unmodified inside AutoGen's \texttt{RoundRobinGroupChat} and reproduces the
in-house ordering exactly, vector ahead on accuracy and efficiency, the full system ahead on
ASR. We therefore claim \emph{transfer} of the memory abstraction and its behavioural
signature, and explicitly not a \emph{gain}: inside AutoGen the full system is $0.088$ F1 behind
vector-only, and its ASR advantage ($5\to4$ attacks) is not significant. We integrated one
external framework, not the two we set out to; DyLAN~\citep{liu2023dylan} is a research
repository rather than an installable package and was not attempted.

\section{Assumptions and reproducibility}
\label{app:repro}

\textbf{Sampling.} $T{=}0.2$, $p{=}0.7$ rather than $1.0/1.0$, to keep sampling noise from
swamping between-configuration effects; output format was verified well-formed at these values.

\textbf{Trust weighting is applied to scores, not vectors.} With L2-normalised vectors under
inner-product search, scaling the score by $\sqrt{\rhotrust}$ is algebraically identical to
rescaling the embedding and avoids re-indexing.

\textbf{One shared read per round.} All agents in a round see the same retrieved context, so
retrieval cost is counted once per round rather than once per agent.

\textbf{Latency} is a serial-equivalent critical path (retrieval plus, per round, the slowest
agent call, plus the judge) computed from per-call latencies recorded at first execution, so it
is stable under cache replay. Wall time is logged but is meaningless once the cache is warm.

\textbf{Tie handling.} Collapse and surface build identical prompts whenever retrieval returned
no conflict, so those pairs are zero by construction: $94/100$ at cap $40$. \texttt{stats.py}
uses \texttt{zero\_method="zsplit"} for Wilcoxon, which keeps them and splits their ranks,
giving $p{=}0.2595$ on the Phase 3 contrast; discarding them gives $p{=}0.0260$ on the same
data. Both are reported in the released \texttt{corpus\_validation.txt}. We rest no claim on
either, since the bootstrap CI on the pre-specified subgroup is the test we declared.

\textbf{Caching.} Every LLM call is disk-cached on the full request payload, so runs replay
byte-identically; we verified this when two formally identical configurations returned F1
agreeing to five decimals. Token counts come from stored server usage records and so do not
depend on cache state. Released: code, per-query JSONL for all 39 runs, the attack set,
\texttt{gates.txt}, \texttt{significance.txt} and \texttt{corpus\_validation.txt}.

\section*{NeurIPS Paper Checklist}

\begin{enumerate}

\item {\bf Claims}
    \item[] Question: Do the main claims made in the abstract and introduction accurately reflect the paper's contributions and scope?
    \item[] Answer: \answerYes{} 
    \item[] Justification: The abstract and \cref{sec:results} state the same numbers: $+0.340$ F1 (CI $[+0.267,+0.414]$) and attacks $22\to7$ of $122$ at $p{=}0.0003$. We also state in the abstract what does \emph{not} hold---two mechanisms cost accuracy---and that our own retrieval cap initially understated the effect eightfold.

\item {\bf Limitations}
    \item[] Question: Does the paper discuss the limitations of the work performed by the authors?
    \item[] Answer: \answerYes{} 
    \item[] Justification: \Cref{sec:limitations} gives five, including the regime split across the retrieval cap, the single model family, and the fact that memory never exceeds the context window on this corpus so retrieval is motivated by cost rather than necessity.

\item {\bf Theory assumptions and proofs}
    \item[] Question: For each theoretical result, does the paper provide the full set of assumptions and a complete (and correct) proof?
    \item[] Answer: \answerNA{} 
    \item[] Justification: The paper makes no theoretical claims; \cref{eq:edge,eq:policy,eq:rho,eq:rank} are definitions of the memory contract and the ranking, not results.

    \item {\bf Experimental result reproducibility}
    \item[] Question: Does the paper fully disclose all the information needed to reproduce the main experimental results of the paper to the extent that it affects the main claims and/or conclusions of the paper (regardless of whether the code and data are provided or not)?
    \item[] Answer: \answerYes{} 
    \item[] Justification: \Cref{sec:setup} and \cref{app:method,app:repro} give the corpora, slices, attack construction, models, sampling parameters, budgets, protocol and statistical procedure. Every LLM call is disk-cached on its full request payload, so a configuration replays byte-identically.

\item {\bf Open access to data and code}
    \item[] Question: Does the paper provide open access to the data and code, with sufficient instructions to faithfully reproduce the main experimental results, as described in supplemental material?
    \item[] Answer: \answerYes{} 
    \item[] Justification: We release the harness, the DebateMem-Conflict attack set, per-query JSONL for all 39 runs, and the scripts that recompute every table (\cref{app:repro}). Both corpora are public.

\item {\bf Experimental setting/details}
    \item[] Question: Does the paper specify all the training and test details (e.g., data splits, hyperparameters, how they were chosen, type of optimizer) necessary to understand the results?
    \item[] Answer: \answerYes{} 
    \item[] Justification: \Cref{sec:setup} specifies slices, item counts, seeds, budget, agent and round counts, sampling parameters, and the retrieval cap; \cref{app:method} gives the trust, ranking and router parameterisation. No parameter was tuned after seeing results---\cref{sec:limitations} states which were fixed a priori and not swept.

\item {\bf Experiment statistical significance}
    \item[] Question: Does the paper report error bars suitably and correctly defined or other appropriate information about the statistical significance of the experiments?
    \item[] Answer: \answerYes{} 
    \item[] Justification: All contrasts are paired on shared items with a $10{,}000$-resample bootstrap CI, a two-sided Wilcoxon test, and exact McNemar on attack discordance (\cref{sec:setup}). We mark an effect established only when the $95\%$ CI excludes zero, and \cref{app:repro} reports the Wilcoxon tie-handling sensitivity explicitly rather than choosing the favourable rule.

\item {\bf Experiments compute resources}
    \item[] Question: For each experiment, does the paper provide sufficient information on the computer resources (type of compute workers, memory, time of execution) needed to reproduce the experiments?
    \item[] Answer: \answerYes{} 
    \item[] Justification: \Cref{app:models} reports measured endpoint throughput for both permitted models and the cost that made the second model family infeasible. Each item is 7 LLM calls; embeddings run on one local GPU.
    
\item {\bf Code of ethics}
    \item[] Question: Does the research conducted in the paper conform, in every respect, with the NeurIPS Code of Ethics \url{https://neurips.cc/public/EthicsGuidelines}?
    \item[] Answer: \answerYes{} 
    \item[] Justification: The research uses public dialogue and QA corpora, involves no human subjects, and releases a synthetic contradiction-injection set whose risks are discussed below.

\item {\bf Broader impacts}
    \item[] Question: Does the paper discuss both potential positive societal impacts and negative societal impacts of the work performed?
    \item[] Answer: \answerYes{} 
    \item[] Justification: \Cref{sec:conflict,sec:gate} show that a memory layer which silently reconciles conflicting facts is exploitable by an adversary who controls timestamps, and that a plausibility-based write filter can make this worse. The positive impact is a cheap mitigation---surface the conflict---requiring no extra model calls.
    
\item {\bf Safeguards}
    \item[] Question: Does the paper describe safeguards that have been put in place for responsible release of data or models that have a high risk for misuse (e.g., pre-trained language models, image generators, or scraped datasets)?
    \item[] Answer: \answerYes{} 
    \item[] Justification: DebateMem-Conflict is template-generated from objects already present in the host corpus, so it produces no novel harmful content and does not transfer outside the corpus it is built from. We release it because the defence we evaluate cannot be checked without it, and we deliberately did not weaken it to flatter our own mechanism (\cref{sec:gate}).

\item {\bf Licenses for existing assets}
    \item[] Question: Are the creators or original owners of assets (e.g., code, data, models), used in the paper, properly credited and are the license and terms of use explicitly mentioned and properly respected?
    \item[] Answer: \answerYes{} 
    \item[] Justification: LoCoMo~\citep{maharana2024locomo}, HotpotQA~\citep{yang2018hotpotqa}, the embedding model~\citep{xiao2024bge} and AutoGen~\citep{wu2023autogen} are cited and used under their public terms.

\item {\bf New assets}
    \item[] Question: Are new assets introduced in the paper well documented and is the documentation provided alongside the assets?
    \item[] Answer: \answerYes{} 
    \item[] Justification: The harness and the attack set are documented in \cref{sec:setup,app:method} and shipped with the scripts that rebuild them deterministically from a seed.

\item {\bf Crowdsourcing and research with human subjects}
    \item[] Question: For crowdsourcing experiments and research with human subjects, does the paper include the full text of instructions given to participants and screenshots, if applicable, as well as details about compensation (if any)? 
    \item[] Answer: \answerNA{} 
    \item[] Justification: No crowdsourcing and no human subjects; all data is from existing public corpora.

\item {\bf Institutional review board (IRB) approvals or equivalent for research with human subjects}
    \item[] Question: Does the paper describe potential risks incurred by study participants, whether such risks were disclosed to the subjects, and whether Institutional Review Board (IRB) approvals (or an equivalent approval/review based on the requirements of your country or institution) were obtained?
    \item[] Answer: \answerNA{} 
    \item[] Justification: No human subjects research was conducted.

\item {\bf Declaration of LLM usage}
    \item[] Question: Does the paper describe the usage of LLMs if it is an important, original, or non-standard component of the core methods in this research? Note that if the LLM is used only for writing, editing, or formatting purposes and does \emph{not} impact the core methodology, scientific rigor, or originality of the research, declaration is not required.
    \item[] Answer: \answerYes{} 
    \item[] Justification: LLMs are the object of study, not a writing aid: they are the debating agents, the judge, and the triple-extraction pass. \Cref{sec:setup} names the model and sampling parameters, and \cref{app:models} reports the measurement that determined the choice.

\end{enumerate}

\end{document}